\documentstyle[11pt]{article}
\addtocounter{section}{-1}
\catcode `\@=11

\if@twoside
\else
\fi
\def\@maketitle{\newpage \null \vskip 2em   % Vertical space above title.
 \begin{center}
  {\bf \@title \par}     % Title set in \bf.
  \vskip 3em                % Vertical space after title.
  {%\large                       % each author set in \large, in a
    \lineskip .5em \sc          % tabular environment
    \begin{tabular}[t]{c}\@author \end{tabular}\par}
  \vskip 2em              % Vertical space after author.
  {\it \lineskip .5em
    \begin{tabular}[t]{c}\@address \end{tabular}\par}
  \vskip 2em              % Vertical space after address.
  {(Received \@date)}           % Date set.
 \end{center}
 \par \vskip 3em}                % Vertical space after date.

\def\address#1{\gdef\@address{#1}}

\def\abstract{\if@twocolumn \section*{Abstract}
\else \small
\begin{center}{\bf Abstract\vspace{-.5em}\vspace{0pt}}\end{center}\quotation
\fi}

\def\endabstract{\if@twocolumn\else\endquotation\fi}

\newcounter{figcaption}
\def\thefigcaption{\arabic{figcaption}}
\def\fnum@figcaption{{\bf Fig. \thefigcaption :}}
\def\figcaption{
  \par\pagebreak {\parindent 0pt \bf Figure Captions} \par \vskip 10pt
  \list{\fnum@figcaption}
  {\leftmargin 5em \labelwidth\leftmargin\advance\labelwidth-\labelsep
   \def\makelabel##1{##1\hfil} \usecounter{figcaption}}
}

\def\thereferences#1{\section*{References\@mkboth
 {REFERENCES}{REFERENCES}}\list
 { \arabic{enumi})\ }{\settowidth\labelwidth{#1)\ }\leftmargin\labelwidth
 \advance\leftmargin\labelsep \usecounter{enumi}}
 \def\newblock{\hskip .11em plus .33em minus -.07em}
 \sloppy \sfcode`\.=1000\relax}

\def\jcite{\@ifnextchar [{\@tempswatrue\@jcitex}{\@tempswafalse\@jcitex[]}}

\def\@jcitex[#1]#2{\if@filesw\immediate\write\@auxout{\string\citation{#2}}\fi
  \def\@citea{}\@jcite{\@for\@citeb:=#2\do
    {\@citea\def\@citea{,\penalty\@m}\@ifundefined
       {b@\@citeb}{{\bf ?}\@warning
       {Citation `\@citeb' on page \thepage \space undefined}}%
\hbox{\csname b@\@citeb\endcsname}}}{#1}}

\newfont{\scrptrm}{cmr8}
\def\@jcite#1#2{${}^{\scrptrm {#1\if@tempswa , #2\fi})}$}

\def\acknowledgement{\if@twocolumn \section*{Acknowledgement}
\else \normalsize \begin{center}
{\bf Acknowledgement\vspace{-.5em}\vspace{0pt}} \end{center}\quotation
\fi}

\def\endacknowledgement{\if@twocolumn\else\endquotation\fi}

\def\@normalsize{\@setsize\normalsize{25pt}\xiipt\@xiipt
\abovedisplayskip 12pt plus3pt minus7pt%
\belowdisplayskip \abovedisplayskip
\abovedisplayshortskip  \z@ plus3pt%
\belowdisplayshortskip  6.5pt plus3.5pt minus3pt}

\def\small{\@setsize\small{22.6pt}\xipt\@xipt
\abovedisplayskip 11pt plus3pt minus6pt%
\belowdisplayskip \abovedisplayskip
\abovedisplayshortskip  \z@ plus3pt%
\belowdisplayshortskip  6.5pt plus3.5pt minus3pt
\def\@listi{\parsep 4.5pt plus 2pt minus 1pt
 \itemsep \parsep \topsep 9pt plus 3pt minus 5pt}}

\def\footnotesize{\@setsize\footnotesize{20pt}\xpt\@xpt
\abovedisplayskip 10pt plus2pt minus5pt%
\belowdisplayskip \abovedisplayskip
\abovedisplayshortskip  \z@ plus3pt%
\belowdisplayshortskip  6pt plus3pt minus3pt
\def\@listi{\topsep 6pt plus 2pt minus 2pt\parsep 3pt plus 2pt minus 1pt
\itemsep \parsep}}

\def\scriptsize{\@setsize\scriptsize{15.8pt}\viiipt\@viiipt}
\def\tiny{\@setsize\tiny{11.6pt}\vipt\@vipt}
\def\large{\@setsize\large{30pt}\xivpt\@xivpt}
\def\Large{\@setsize\Large{36.6pt}\xviipt\@xviipt}
\def\LARGE{\@setsize\LARGE{41.6pt}\xxpt\@xxpt}
\def\huge{\@setsize\huge{50pt}\xxvpt\@xxvpt}

\def\section{\@startsection {section}{1}{\z@}{-3.5ex plus -1ex minus
    -.2ex}{2.3ex plus .2ex}{\normalsize\bf}} %mod%
\def\subsection{\@startsection{subsection}{2}{\z@}{-3.25ex plus -1ex minus
   -.2ex}{1.5ex plus .2ex}{\normalsize\it}}  %mod%
\def\subsubsection{\@startsection{subsubsection}{3}{\z@}{-3.25ex plus
 -1ex minus -.2ex}{1.5ex plus .2ex}{\normalsize\it}} %mod%

\def\thesection{\S\arabic{section}.}
\def\thesubsection{\arabic{section}.\arabic{subsection}}

\def\appendix{\par
  \@appendix  %add%
  \setcounter{section}{0} \setcounter{subsection}{0}
  \def\thesection{\Alph{section}.}
  \def\thesubsection{\thesection\arabic{subsection}} %mod%
  \@addtoreset{equation}{section}   % Makes \section reset 'equation' counter.
  \def\theequation{\Alph{section}.\arabic{equation}} %mod%
}

\def\@appendix{{\addvspace{3ex} \begin{center}{\bf APPENDIX}\end{center}}}

\catcode `\@=12

\renewcommand{\thesection}{{\bf\S}\arabic{section}}
\newcounter{tmpequation}

   \def\theequation{\arabic{section}.\arabic{equation}}
\def\tmptheequation{\arabic{section}.\arabic{tmpequation}}

\def\eqnal{
    \addtocounter{equation}{1}
	\setcounter{tmpequation}{\value{equation}}
	\setcounter{equation}{0}

	\let\savetheequation=\theequation
	\renewcommand{\theequation}{\tmptheequation\alph{equation}}
}
\def\endeqnal{
	\setcounter{equation}{\value{tmpequation}}
	\let\theequation=\savetheequation
}

\newenvironment{eqnumalpha}{\eqnal}{\endeqnal}

\newcommand{\bequ}{ \begin{equation} }
\newcommand{\eequ}{ \end{equation} }
\newcommand{\barr}{ \begin{array} }
\newcommand{\earr}{ \end{array} }
\newcommand{\beqarr}{ \begin{eqnarray} }
\newcommand{\eeqarr}{ \end{eqnarray} }

\newcommand{\baralpha}{ \begin{eqnumalpha} \beqarr}
\newcommand{\earalpha}{ \eeqarr \end{eqnumalpha}}

\def\I{{\rm I}}

\def\Del#1{\Delta_{#1}}

\def\al {\alpha}
\def\vf{v_{F}}
\def\omeg0{\omega_{0}}
\def\omga{\omega_{\alpha}}

\def\Omeg0{\Omega_0}

\def\d{{\rm d}}
\def\sgn{{\rm sgn}}
\def\i{{\rm i}}

\def\m1{m_{1}}
\def\d{{\rm d}}

\def\i{{\rm i}}

\def\sgn{\rm sgn }

\begin{document}
%\draft
%\tighten
%
%\twocolumn[

\title{ {\bf  Isotope Effect on the Formation Energy of Soliton
             in t-Polyacetylene }   }

 \author{           Ry\=oen S{\sc hirasaki}}
 
\address{{ Department of Physics, Faculty of Education } \\
{ Yokohama National University,
 Yokohama 240}
 }

\date{\today}

\maketitle

%\vspace{1cm}
%\widetext

\begin{center}

{\large\bf Abstract}

\end{center}

 The formation energy of a soliton was derived in the TLM model to be
\( 2\Delta_{0}/\pi \) which is independent of the masses of the polymer atoms.
We estimate effects of the electron-phonon coupling on the soliton formation
energy, particularly, through modifications of the harmonic oscillations of
(CH) groups.
A model based on the amplitude mode formalism is used and the electron-phonon
coupling is taken into account to make the effective potential of lattice
vibration. Among dynamical degrees of freedom in the
lattice vibration, a translational mode of the soliton is separately examined,
since it corresponds to a zero-energy mode.
We get a difference in the formation energy between the two isotopes of about
0.03eV. Soliton trapping around  defects or impurities changes the formation
energy and enlarges the isotope difference.
%Finally, the difference becomes the magnitude of 0.03eV

%\end{abstract}
%\pacs{}
%\narrowtext
%]

%\newpage

\section{Introduction}

 The trans-polyacetylene is an ideal material which has excitations such as
solitons, polaron, bipolaron, and so on. They can be described very well by the
SSH model which was proposed by  Su {\it et al}.\jcite{Su} This model contains
the electron phonon interaction in addition to electron hopping and lattice
vibrations.
Takayama {\it et al.}\jcite{Takayama} made the TLM model which is a continuum
version of the SSH model.
It makes the analytical study of polyacetylene possible.

 A number of experimental studies have supported the presence of a kink
excitation in t-polyacetylene (t-(CH)$_x$).$^{3)-6)} $ Suzuki {\it et al.} have
reported an optical absorption measurement in lightly doped
polyacetylene.\jcite{Suzuki} They have found that the absorption related to the
electronic mid-gap state is enhanced by doping. This result has been
interpreted as transitions between the mid-gap levels of solitons,
induced by the doping, and the conduction band.

In addition to the absorption due to the mid-gap scattering transition, sevral
absorptions have been observed near $1370 {\rm cm}^{-1}$ and $900 {\rm
cm}^{-1}$ in the experiments on the infra-red absorption for a lightly doped
(CH)$_x$.\jcite{Fincher} Mele and Rice have suggested that these peaks are due
to lattice vibration.\jcite{Mele1}
 Horovitz has explained these peaks as the lattice vibration around a soliton
added to a dimerized state, proposing a model with several phonon bands. This
model is called "the amplitude mode formalism".\jcite{Horovitz}

  On the other hand, the experimental studies of the photo-induced absorption
spectra of undoped t-(CH)$_x$ have also shown additional infra-red active
modes.{\jcite{Var2}\jcite{Tanaka}} The positions are listed in Table I.
 These infra-red active peaks have been considered to be due to  internal
vibrational modes created by the presence of the soliton. The intensities of
these new modes are much weaker than those of the first three modes.

%These IR active modes have been observed by photoinduced absorption
%%experiments%. In this method, pumping photons are used to induce
%%electron-hole pairs and IR% probe photons are used for observations.
%The pumping photons excite electrons from the valence band to the conduction
%%ba%nd. The created electron-hole pair  decays into a soliton-antisoliton pair
%%or t%wo polarons in a short time.
%The second photon is absorbed when the frequency of the photon coincides with
%%t%hat of vibration modes around the solitons, antisolitons, and polarons.
%The relative intensity of the IR absorption increases as $\sqrt{I}$ where $I$
%%i%s the intensity of pumping photons.

Ito {\it et al.} have studied the phonon structures in the presence of a
soliton in the framework of the continuum theory for trans-polyacetylene (TLM
model).
 They have analyzed small lattice oscillations by using time-dependent
adiabatic mean-field equations resulting from a selfconsistent treatment of the
electronic field and the phonon field.\jcite{Ito} They have found three
localized modes: the Goldstone mode (or the translational mode), the amplitude
oscillation mode (the shape mode) and a third mode.
Ito {\it et al.} have also found that two of the three localized modes, i.e.
the Goldstone mode (G.M.) and the third localized mode are IR active, whereas
the amplitude oscillation mode is inactive.\jcite{IA}

Mele and Hicks have applied Horovitz's amplitude mode formalism to investigate
the optical properties of the localized phonon modes.\jcite{Mele}
They have shown that there are three absorption peaks due to the translational
mode in the presence of a single soliton.
 Terai, Ono, and Wada have applied the Mele-Hicks theory to calculate the
phonon structures, taking account of the translation and the third localized
modes, where they have used three phonon bands.\jcite{Terai}
(For each lattice site, there are several degrees of freedom such as bond
vibration between C and H and that between nearest neighboring C atoms. Within
the polymer plane, there are 6 degrees of freedom.\jcite{Schugerl}
 The number of IR-active modes e.g. gerade modes is, explicitly, 4.)
 Heeger {\it et al.} performed an optical-absorption experiment with high
resolution, observing two addtional IR peaks.\jcite{Schaffer}
Terai {\it et al.} have estimated relative optical absorption intensities and
obtained a good agreement with the experiment.\jcite{Terai1989}

In this paper, we would like to study an isotope effect on soliton formation
energy in t-polyacetylene.
attice oscillations.\jcite{Heeger}

 of the unharmonicity of the effective potential gives a correction to the
phonon frequency and  the electron-phonon coupling constant. We will show that
the correction can give the difference in the formation energy as large as the
observed value.

\section{The effective potential of the one-dimensional
electron-phonon system}

\label{I.1}

 In quasi-one-dimensional electron phonon system bearing the Peierls
transition, there are two types of phonon modes.
 One is the amplitude mode and the other is the phase mode (the CDW sliding
mode).
% However in the half filling case, the effect of the sliding mode is small in
%%comparison with the amplitude mode.
In the incommensurate charge density wave, the ion displacement at the $m$th
site is
$$
u_{m}=u_0 \cos(q_0\ ma +\delta)
$$
where  $u_0$ is the displacement amplitude, $q_0$ is the wave number which is
equal to $2k_F$. The phase variable $\delta$ describes the motion of the charge
density wave.
 In the half-filling case such as trans-polyacetylene, $q_0=\pi/a$, so that
$u_m =(-)^m u_0 \cos \delta$.
Thus the dynamics of $\delta$ in t-polyacetylene is written by using only the
amplitude of the charge density wave.

 Let us consider a one-dimensional electron system including a electron-phonon
coupling with a optical phonon field $\Delta=g u_0$ where g is the
electron-phonon coupling constant. the Hamiltonian is given by
\beqarr
H & = &  \frac{1}{2\pi v_F \lambda \omega_Q^2} \int \d x
[\dot{\Delta}^2(x)+\omega_Q^2\Delta^2 (x) ] \nonumber \\
 & & \nonumber \\
 & & + \sum_s\int\d x \psi_s^\dagger (x)[-\i v_F\sigma_3\partial_x
+ \sigma_1 \Delta(x)] \psi_s(x),
\label{1.1}
\eeqarr
where $\omega_Q$ is the bare phonon frequency. The electron field $\psi_s$ is
introduced as a two component spinol field representing left and right-going
waves at the Fermi level.
 $\lambda$ is a dimensionless electron-phonon coupling constant.
the Hamiltonian (\ref{1.1}) is called the Takayama-Lin-Liu-Maki (TLM) model.
\jcite{Takayama}

 The coupling between electron and phonon fields induces unharmonicity of the
lattice vibration and, due to this non-linearity, the Hamiltonian (\ref{1.1})
bears a bond order wave state (BOW state).

To investigate the unharmonicity, we make a effective potential for the phonon
field using loop expansion of a partition function.

The partition function of (\ref{1.1}) is written as follows:
\bequ
Z=Z_0 \left< T_\tau exp\left( -\int_0^\beta H_I \d \tau \right) \right>.
\label{1.2}
\eequ
where
$\left< T_\tau \right>$ being a thermal average with respect to the unperturbed
Hamiltonian. $H_I$ is the perturbation term given by the electron-phonon
coupling $\psi_s^\dagger \sigma_1 \Delta \psi_s$. $Z_0$ is the partition
function of the unperturbed Hamiltonian in which the electron-phonon coupling
is excluded.
$Z_0$ is written in the path integration form,
\beqarr
 Z_0 & = & \int(D\Delta D\psi_s^\dagger D\psi_s)
\exp\biggl\{
-\int\d \tau (L_\Delta^0+L_\psi^0)\biggr\}
\nonumber \\
 & & \nonumber \\
 L_\Delta^0 & = & \frac{1}{2\pi v_F \lambda {\omega_Q}^2 }
\int \d x \Biggl[
\Biggl( \frac{\partial\Delta}{\partial \tau} \Biggr)^2
+{\omega_Q}^2 \Delta^2(x)\Biggr]
\nonumber \\
 & & \nonumber \\
 L_\psi^0 & = & \int\d x \psi_s^\dagger (x)
\frac{\partial \psi_s (x)}{\partial \tau}
 + \sum_s\int\d x \psi_s^\dagger (x)[-\i v_F\sigma_3\partial_x
%
%+ \sigma_1 \Delta(x)
%
] \psi_s(x),
\label{h0}
\eeqarr
The thermal average in (\ref{1.2}) is also written by the path integration
form.
The path integration over $\psi_s$ is approximated by a diagramatic
calculation.
%
%\bequ
%Z=\int D \Delta e^{-\int \d \tau H_0 -\int \d \tau \d x U}
%\label{1.2'}
%\eequ
%
 Consider the summation over connected loop diagrams which are shown in
Fig.1.\jcite{Luttinger}\jcite{CW} The solid line is associated with the
electron Green function. The solid circle corresponds to the vertex with
ˆŸ$\sigma_1$ coupling.
In the momentum space,  the electron Green function is given by
\beqarr
G^{(0)}(\omega,k) \delta_{s,s'} & = &
-\int\int \d x\d \tau\left< T_\tau \psi_s(x,\tau)\psi^\dagger
_{s'}(x',0)\right> e^{-\i k x -\i \omega\tau}
\nonumber \\
 & = & \frac{\i \omega \I + \vf k\sigma_3}{ (\i \omega )^2 - (\vf k)^2}
\delta_{s,s'} .
\label{1.2"}
 \eeqarr
where $s$ denotes up or down of the spin orientation.  Representing the sum of
the diagrams in Fig.1 as $U$, the partition function is approximately given by
\bequ
 \frac{Z}{Z_0}  =  \frac{
  \int (D\Delta) \exp
\Biggl\{
-\int\d\tau L_\Delta^0+\int\d \tau d x U
\Biggr\} }
         {   \int (D\Delta) \exp
\Biggl\{
-\int\d\tau L_\Delta^0
\Biggr\}  }  .
\label{1.2'}
\eequ
 Let us estimate $U$. The lowest order term with respect to the momentum
transfer $q$ between the electron, $\psi_s$, and the lattice, $\Delta(x)$, is
given by the summation written as follows:
\beqarr
 &  & -\frac{1}{\beta} Tr \sum_{n,s,\omega}\int\d k
\frac{1}{4\pi n}
  [ \Delta(0)\sigma_1 G^{(0)}(\i\omega,k)] ^n
 \nonumber \\
  & = & \frac{1}{\pi\beta}\sum_{ \omega } \int\d k \ln
\left(1-\frac{\Delta(0)^2}{(\i \omega)^2-(\vf k)^2 }\right) ,
\label{1.3}
\eeqarr
where the summation over $\omega$ is taken over the Matsubara frequency
$\omega=T(2m+1)\pi$,  $m$ being an integer. $\Delta(0)$ is the Fourier
transform of $\Delta(x)$ at $q=0$.
 Taking zero temperature limit and using the relation
\beqarr
\frac{1}{\beta}\sum_{\omega,\mbox{odd}} \ln
\left(\frac{\i\omega-|a|}{\i \omega-|b| } \right)
 & = &
\frac{1}{4\pi\i}\int\d \varepsilon \tanh
\frac{\beta\varepsilon}{2}\ln\frac{\varepsilon-|a|}{\varepsilon-|b|}
 \nonumber \\
 & = & \frac{1}{2}(|a|-|b|), \quad (T\rightarrow 0).
\eeqarr
  we can rewrite (\ref{1.3}) as
\bequ
 - \frac{1}{\pi\vf}\Delta^2 (0)
\left( \ln\frac{\Delta(0)}{2\Lambda} -\frac{1}{2}
\right),
\label{1.4}
\eequ
where $\Lambda$ is the electron band cut off given by $\Lambda=\vf /a$.

The higher order terms with respect to $q$ is given by diagrams shown in Fig.2.
 The summation over these diagrams gives
\beqarr
 & & -\frac{1 }{\beta} Tr \sum_{\omega,s}
\int \frac{\d k}{4\pi}
\left[
\left(
\Delta \sigma_1
\frac{\i \omega+\vf (k+q) \sigma_3 + \Delta\sigma_1}{(\i\omega)^2-\vf^2
(k+q)^2-\Delta^2}
\right)
\left( \Delta \sigma_1
\frac{\i \omega+\vf k \sigma_3 + \Delta\sigma_1}{(\i\omega)^2-(\vf
k)^2-\Delta^2}
\right) \right.
  \nonumber \\
 & &\nonumber \\
 & & \left.  -
\left(
\Delta \sigma_1
\frac{\i \omega+\vf k \sigma_3 + \Delta\sigma_1}{(\i\omega)^2-(\vf
k)^2-\Delta^2}
\right) ^2
\right]
\nonumber \\
 & &\nonumber \\
 & & =  -\frac{1}{\pi\vf}Ê\Delta \left[
\frac{ \sqrt{ 4 \Delta^2(0)+(\vf q)^2 } }{\vf q}
\ln \left( \frac{\vf q+\sqrt{4\Delta^2(0) +(\vf q)^2}}{2 \Delta(0)} \right)
-1
\right] \Delta.
\label{1.5}
\eeqarr
where we subtracted from the first term in the right hand side of (\ref{1.5})
the $q$ independent term (the second term).
%We neglected frequency dependence of the diagrams.
%
In the real space, this term would be expressed as
\bequ
 - \frac{1}{\pi\vf} \Delta(x) \left(\frac{  \sqrt{ 1+\gamma^2 }  }
{\gamma}\ln (\gamma+\sqrt{1+\gamma^2})-1
\right)\Delta(x)
\label{1.6}
\eequ
where
 \bequ
\gamma = -\frac{\vf}{2\Delta(0)}\i \frac{\partial}{\partial x} .
\label{1.6'}
\eequ
 $\Delta(0)$ is considered to be absolute of the mean value of $\Delta(x)$.
Since $U$ is expressed by the sum of (\ref{1.4}) and (\ref{1.5}), we can study
the lattice dynamics of the electron-phonon system (\ref{1.1}) by using the
total effective potential which is given as follows :
\bequ
\frac{1}{2\pi\vf } \int\d x
\left[ \frac{\Delta(x)^2}{\lambda}+
 2 \Delta(x)\left(\frac{  \sqrt{ 1+\gamma^2 }  }
{\gamma}\ln (\gamma+\sqrt{1+\gamma^2})-1
\right) \Delta(x)
+2 \Delta(x)^2 \ln\frac{|\Delta(x)|}{2\Lambda}   -\Delta(x)^2
\right].
\label{1.7}
\eequ
In the perfectly dimerized state, and in the adiabatic limit, the spatial
variance of $\Delta(x)$ is equal to zero.
 The effective potential have double minimum points at $\Delta=\pm \Delta_0$
where $\Delta_0$ is determined by
\bequ
\Delta_0=2\Lambda\exp \left\{-\frac{1}{2\lambda} \right\}.
\label{1.7'}
\eequ
  The potential is written by the dashed line in Fig.3. Thus, we consider  that
$|\Delta(0)|$, which was introduced in (\ref{1.6}) as
 the mean value of $\Delta(x)$, is given by $\Delta_0$. Therefore,  the
effective Hamiltonian is given by
\beqarr
H & = & \frac{1}{2\pi\vf \lambda\omega_Q^2}\int \d x \left[
\dot{\Delta}(x)^2+\omega_Q^2\Delta(x)^2\right]
\nonumber \\
  & &\nonumber \\
 &  & + \frac{1}{2\pi\vf} \int\d x \left[
 2 \Delta(x) D_\gamma
 %\left(\frac{ \sqrt{ 1+\gamma^2 } }
%{\gamma}\ln (\gamma+\sqrt{1+\gamma^2})-1
%\right)
\Delta(x)
+ 2 \Delta(x)^2
\left( \ln \frac{|\Delta(x)|}{2\Lambda} - \frac{1}{2}
\right) \right],
\nonumber \\
 & & \label{1.8}
\eeqarr
where $\dot{\Delta}(x)$ is the time-derivative of $\Delta(x)$. $D_\gamma$  and
$\gamma$ are introduced as follows :
\bequ
D_\gamma=\frac{ \sqrt{1+\gamma^2} }{\gamma} \ln (\gamma+\sqrt{1+\gamma^2})-1,
\label{di}
\eequ
\bequ
\gamma=-\frac{\xi_0}{2}\frac{\partial}{\partial x},
\label{ga1}
\eequ
where $\xi_0=v_F/\Del0$.  As is well-known, the electron-phonon system
(\ref{1.1}) bears the non-linear excitation. A soliton is the solution of
(\ref{1.1}) which seperates $\Delta(x)=\Del0$ phase and $\Delta(x)=-\Del0$
phase.
 The effective Hamiltonian (\ref{1.8}) also has a soliton solution. It's
spatial form is shown in Fig.4.
We obtained this form by using the variational method.
(Appendix A.) The form of $\Delta(x)$ for the soliton is  nearly equal to  the
well-known form.
\bequ
\Delta_s(x)=\Delta_0\tanh  x/\xi,
\label{Ds}
\eequ
which is the solution of the Hamiltonian (\ref{1.1}).\jcite{Takayama}
 Our result gives a value of the soliton width of about $\xi=1.04\xi_0$ with
$\xi_0=v_F/\Del0$.

The soliton excitation energy which is centered around the soliton is mainly
kept in the electron system. In a soliton case, the energy level of the
electronic states in the valence band are shifted by the lattice distortion and
the mid-gap localized state is induced.\jcite{Takayama}  In our model, the
property of the electron system is approximated by the effective potential
which is constructed on the assumption of the continuum electronic dispersion.
Since the localized electronic state is not included,
 our model, (\ref{1.8}), does not include the exact change of the electronic
energy level.
 Using (\ref{Ds}), and using the fact that $\Delta_s(x)$ is the solution of
(\ref{a1}), the soliton formation energy is derived to be $\Del0/\pi$ (Appendix
A).
 Our result gives the different value with the result given by the TLM
mode.l\jcite{Takayama}
  On the other hand,  the Hamiltonian (\ref{1.8}) gives good coincidence with
the TLM model, (\ref{1.1}), in analyzing the lattice dynamics of polymer chain.
 As we will show later, the dispersion relation of phonon and the localized
ocillation mode around the soliton obtained in the TLM model are well
reproduced in our model. Then, the Hamiltonian (\ref{1.8}) would be suitable
for analyzing the renormalization of the unharmonic lattice oscillation and for
making the correction of the parameters $\lambda$, $\omega_Q$, etc.

 The eigen mode $g_\ell$ of the lattice fluctuation $\eta$ around a  static
soliton is given by the Schr\"odinger equation
\bequ
 - \frac{1}{ 2\lambda\omega_Q^2 }\ddot{g}_\ell=
\left[ \frac{\sqrt{1+\gamma^2}}{\gamma}
\ln (\gamma + \sqrt{1+\gamma^2}) - 1 \right]
 g_\ell
 + \left( 1+\ln\frac{|\Delta_s(x)|}{\Del0}\right) g_\ell  ,
\label{1.9}
\eequ
where $\Delta_s(x)$ is the static soliton solution of (\ref{1.8}), {\it i.e.}
(\ref{Ds}).  In the perfectly dimerized state,  the $\log$ term of  the
potential in (\ref{1.9}) vanishes and the continuum spectral of phonon
frequency $\Omega_q$ is given by
\bequ
 \frac{\Omega_q^2}{ 2 \lambda\omega_Q^2}
=\frac{ \sqrt{1+s^2} }{ s }\sinh^{-1} s,
 \label{1.10'}
\eequ
 where $s=\vf q/(2\Del0)$, $q$ being the wave number of the phonon. Then, the
extended modes of lattice fluctuation around the  soliton have the spectrum
like (\ref{1.10'}).
 This dispersion relation was also derived by Nakahara and Maki  using the
random phase approximation (RPA) method for the  electron
polarization.\jcite{N.M}

 In the Hamiltonian (\ref{1.1}), there are localized modes  which have non-zero
amplitude around the soliton center. They are the Goldstone mode, amplitude
oscillation mode and the third localized mode of soliton. These localized modes
have been obtained by several groups using the RPA
method.\jcite{Ito}$^,$\jcite{N.M}$^-$\jcite{Hicks} We show in Appendix B that
(\ref{1.9}) has the zero-energy mode (the Goldstone mode) and that its form is
given by $\frac{\d\Delta_s(x)}{\d x}$.
Substituting the form of the soliton which is shown in Fig.4
into (\ref{1.9}),
the frequencies of the other localized modes are numerically obtained.
(Appendix B)
 Their frequencies are shown  in Table II. Our result coincides with other
works.\jcite{Ito}\jcite{LM}\jcite{Hicks}

 In this paper, we discuss the quantum correction to the soliton formation
energy using (\ref{1.8})
 and soliton solution (\ref{Ds}).

\section{Anharmonicity of the lattice vibration}

Because the potential of the lattice field $\Delta$ has the logarithmic term,
the lattice fluctuation $\eta$ around $\Delta=\pm \Delta_0$ has an
anharmonic potential.
In (\ref{1.1}), the parameters $\omega_Q$, $\lambda$ and $\vf$ were determined
so as to obtain good agreements with experimental values of the electron band
gap and the phonon dispersion considering the lattice fluctuation as the
harmonic lattice oscillation.
  In the dispersion relation (\ref{1.10'}) and the selfconsistency equation for
the electron band gap, (\ref{1.7'}), these parameters were used.
 However, since anharmonicity of the lattice oscillation modifies the mean
position of (CH) groups and the energy spectral of phonon dispersion, we should
make corrections for the parameters $\lambda$, $\omega_Q$ and $\vf$.  In the
perfectly dimerized case, expanding $\Delta(x)$ around $\Delta=\Delta_0$, the
Hamiltonian becomes
\beqarr
H & \simeq & \frac{1}{2\pi\vf \lambda}
\int\d x \left(
\frac{\dot{\eta}^2}{\omega_Q^2} +
2\lambda
\eta D_\gamma
%\left(\frac{ \sqrt{1+\gamma^2} }
%{\gamma}\ln (\gamma+\sqrt{1+\gamma^2})-1
%\right)
\eta
+2\lambda\eta^2+
\frac{2\lambda}{3\Delta_0}\eta^3-\frac{\lambda}{6\Delta_0^2}\eta^4
\right)
\nonumber \\
 & & -\frac{1}{2\pi\vf}\int \d x \Delta_0^2 .
\label{2.1}
\eeqarr
where $\Delta(x)=\Delta_0+\eta$. Suppose that the higher order terms in
$\eta/{\Del0}$ are sufficiently small in the quasi-one dimensional system,
we neglected  terms over $5$th order with respect to  $\eta/\Delta_0$.  We
compare the potential of (\ref{1.8}) and (\ref{2.1}) graphically in Fig.3. The
potential of (\ref{2.1}) is drawn by the solid line. The coincidence is fairly
well when $|\eta/\Delta_0|<0.5$.
 For the realistic material, t-polyacetylene, the parameters are
$\lambda\sim0.2$, $\omega_Q\sim0.3eV$, and $\Delta_0\sim0.67$eV,
then
\bequ
\frac{\left< \eta^2 \right>}{\Delta_0^2} =
\left<
 {\rm T}_\tau \eta(x,\tau=0^+)\eta(x,0)
\right>
=\frac{1}{4\Delta_0^2}
\int\d (\vf q)
\frac{\lambda\omega_Q^2}{\omega_q} \simeq 0.4 ,
\label{2.2}
\eequ
 where $\omega_q$ is given by (\ref{1.10'}).  Considering that over the 5th
order contribution from $\eta/\Del0$ is negligible, we investigate the effect
of lattice fluctuation described by (\ref{2.1}).
The bare phonon Hamiltonian which is written by the first three terms in the
right hand side of (\ref{2.1}) gives the bare Green function of $\eta$,
\beqarr
D_\eta (\omega,k) & = &- \int \left< T_\tau \eta(x,\tau) \eta(x',0)\right>
e^{-\i k x}e^{-\i\omega\tau}
\nonumber \\
 & = & \frac{\pi\vf\lambda\omega_Q^2}{(\i\omega)^2-\Omega_k^2} .
\label{2.3}
\eeqarr
 The  forth and the fifth terms in the right hand side of (\ref{2.1}) give 3
point and 4 point vertices, respectively.
 They are associated with the diagrams in Fig.5.

\section {Correction to the soliton formation energy}

The 3 point and the 4 point vertices give the correction to spectral of the
harmonic oscillation. We will make one-loop approximation and estimate the
correction to the soliton formation energy. In the one loop approximation, the
3 point vertex gives the effective contribution which is shown by  diagrams in
Fig.6a).
\bequ
 \int \d x \eta \cdot \frac{1}{2}\cdot\frac{2}{\pi\vf\Delta_0}\left< \eta^2
\right>=
\int\d x \frac{\Delta_0}{\pi\vf}\delta m\ \eta.
\label{2.4}
\eequ
In the same way, the 4 point vertex gives the effective
 contribution which is shown in Fig.6b).
\bequ
- \int\d x \eta^2 \frac{1}{2\pi\vf}\delta m .
\label{2.5}
\eequ
Here, we introduced the notation $\delta m=\left<\eta^2\right>/\Delta_0^2$.
Totally, the contribution becomes
\beqarr
 & & \frac{\delta m}{2\pi\vf} \int\d x (-2\eta\Delta_0+\eta^2)
\nonumber \\
 & \simeq & -\frac{\delta m}{2\pi\vf}
\int \d x \left( \left(\frac{1}{\lambda}-1\right)\Delta(x)^2+2\Delta(x)^2 \ln
\frac{|\Delta(x)|}{2\Lambda}
\right)
\nonumber \\
 & &  -\frac{\delta m}{2\pi\vf} \int \d x \Delta_0^2.
\label{2.6}
\eeqarr
Since $\eta=\Delta(x)-\Delta_0$ is a small quantity, and since the functional
form of the correction term should be similar to the effective potential in
(\ref{1.9}), we approximated the first line of (\ref{2.6}) by the second line.

Because the parameters $\lambda$, $\omega_Q$, and $\vf$ in (\ref{1.1}) is
selected to give the good coincidence with the realistic material in the
mean-field theory, the Hamiltonian should be  reconstructed to cancel the
contribution (\ref{2.6}). As the corrected  Hamiltonian, we obtain
\beqarr
H_{R}'& = & \frac{1}{2\pi\vf\lambda\omega_Q^2}\int\d x
(\dot{\Delta}(x)^2+\omega_Q^2 \Delta^2(x))
\nonumber \\
 & &\nonumber \\
 & & \frac{1}{2\pi\vf}\int\d x
\left[
2 \Delta(x) D_\gamma
%\left(\frac{  \sqrt{ 1+\gamma^2 }  }
%{\gamma}\ln (\gamma+\sqrt{1+\gamma^2})-1
%\right)
\Delta(x)
+ 2\Delta(x)^2 \ln \frac{|\Delta(x)|}{2\Lambda}
-\Delta^2(x)
\right]
\nonumber \\
& & \nonumber \\
& &  + \frac{\delta m}{2\pi\vf} \int\d x \left[
\left(\frac{1}{\lambda}-1\right) \Delta(x)^2
+ 2\Delta(x)^2 \ln \frac{|\Delta(x)|}{2\Lambda}
-\Delta^2(x)
\right]
\nonumber \\
& & \nonumber \\
 & & +\frac{\delta m }{ 2\pi \vf}\Delta_0^2 L .
\label{2.7}
\eeqarr
where $\gamma =-\i \frac{\vf}{2\Del0}\frac{\partial}{\partial x}$.  The third
and the forth line in (\ref{2.7}) is called as the "counter  term" which is
added to cancel the contribution (\ref{2.6}). In the derivation of $H_{R}'$, we
assumed that the fluctuation $\eta/\Del0$ is smaller than $1$, and  terms over
the 4th order with respect to $\eta/\Del0$ is negligible. In the quasi
one-dimensional  electron-phonon system, like polyacetylene, this condition
would be satisfied.
 In the purely one-dimensional system, the higher order terms with respect to
$\eta/\Del0$ is not negligible.  Then, we can not renormalize the anharmonicity
into the form of (\ref{2.7}). We do not consider the purely one-dimensional
case.

Next, we renormalize the unharmonic contributions which come from the counter
terms in (\ref{2.7}). Repeating these procedures, we obtain the renormalized
Hamiltonian
% (\ref{2.7}) is the Hamiltonian
 which represents the actual situation of the quasi one-dimensional
electron-phonon system at the half-filling. By the redefinition of the
parameters $\lambda$, $\vf$ and $\omega_Q$, the renormalized Hamiltonian is
written in the form of (\ref{1.8}). Comparing (\ref{1.8}) and (\ref{2.7}), we
obtain the renormalized Hamiltonian
\beqarr
H_R & = & \frac{1}{ 2\pi{\vf}_R \lambda_R {{\omega_Q}_R}^2 }
\int \d x \left[ \dot{\Delta}_R(x)^2+  {{\omega_Q}_R}^2 \Delta_R (x)^2\right]
\nonumber \\
  & & \nonumber \\
&  & + \frac{1}{2\pi {\vf}_R }
\int\d x \left[
2 d_R \Delta_R(x)  D_\gamma
%\left(\frac{  \sqrt{ 1+\gamma^2 }  }
%{\gamma}\ln (\gamma+\sqrt{1+\gamma^2})-1
%\right)
\Delta_R(x)
 \right.
\nonumber \\
 & & \nonumber \\
& & \left. + 2 \Delta_R(x)^2
\left(\ln \frac{|\Delta_R(x)|}{2\Lambda_R}
-\frac{1}{2} \right) \right].
\label{2.7'}
\eeqarr
\beqarr
 {\omega_Q}_R & = & (1-\delta m)^{-1/2} (1+\lambda\ln (1-\delta m) )^{1/2}
\omega_Q
\nonumber \\
  \lambda_R & = &
 (1+\lambda\ln (1-\delta m) )^{-1} \lambda
\nonumber \\
\Delta_R & = & (1-\delta m)^{-1/2} \Delta
\nonumber \\
 d_R & = & (1-\delta m)
\nonumber \\
 {\vf}_R & = &  \vf
\nonumber \\
 \Lambda_R & = & {\vf}_R/a
\label{2.7"}
\eeqarr
where $a$ is the lattice constant. We determined (\ref{2.7"}) on the condition
that ${\vf}_R$ and $\Lambda_R$ are invariant under the renormalization since
they are parameters of the electron system.
Because $\delta m >0$, the corrected parameters ${\omega_Q}_R$ and $
{\lambda}_R $ are larger than the bare values.
{}From $\frac{\Delta_R^2}{2\pi {\vf}_R }=(1-\delta m)^{-1}\frac{\Delta^2}{2\pi
{\vf} }$,  we can see that, in the actual material, the effective potential is
$(1-\delta m)^{-1}$ times deeper than the uncorrected one.

On the other hand, since the effective potential becomes deeper, the
differential equation for the soliton is affected by the renormalization. We
calculated the soliton width $\xi_R$ for the various values of $1/d_R=(1-\delta
m)^{-1}$ using the variational method. (Appendix A)
We plot $\xi_0/\xi_R$ in Fig.7 as the function of $1/d_R$. The soliton width
becomes narrower as the effective potential becomes deeper. It behaves like
\bequ
\frac{\xi_R}{\xi_0}=1.04 {d_R}^{0.12} =1.04 (1-\delta m)^{0.12} ,
\label{2.7s}
\eequ
in the range $0.2<{d_R}^{-1}<2$.

 In the static limit, $\dot{\Delta}=0$, using (\ref{a1}) , the soliton
formation energy is given by
\beqarr
 & & \frac{1}{2\pi {\vf}_R} \int\d x ( {\Delta_0}_R^2 - \Delta_R(x)^2)
 \nonumber \\
% & &
%+2\left(\Delta(x)^2\ln\frac{|\Delta(x)|}{2\Lambda}
%- \Del0^2\ln\frac{|\Delta_0|}{2\Lambda}
%\right)
%\right]
%\nonumber \\
& & \simeq \frac{\xi_R}{\pi \xi_0 } \frac{ {{\Del0}_R}^2 }{\Del0}
\label{2.8}
\eeqarr
 where, from the first to the second line, we used the soliton form $
\Delta(x)=\Del0 \tanh( x/\xi_R ) $.
%The value $1/\pi$ is not independent of coupling constant.
 In addition to (\ref{2.8}), zero-point lattice oscillation gives a
contribution to the total energy, which is explained by the path
integration.\jcite{CW}  The presence of the soliton changes the phonon
frequency and this contributes to the soliton formation energy.  Then,  the
soliton formation energy is, totally, given by
\bequ
 E_s = \sum_\ell \frac{1}{2}\Omega_\ell\biggl|_{\mbox{soliton}}
 - \sum_\ell \frac{1}{2}\Omega_\ell\biggl|_{\mbox{dimer}}
+ \frac{1 }{(1-\delta m)^{0.88} }
 \frac{\Del0}{\pi} .
\label{2.9}
\eequ

\section{The effect of impurity}

 In the realistic material, there are lattice defects or impurities.
The potential due to the impurities traps the soliton and therefore the
formation energy is affected by the trapping energy.
 Suppose  spin or charge of the soliton interacts with impurity and the
impurity potential is given by the delta function $-V\delta (x)$.
If the soliton is rigid, the trapping energy will be
\bequ
 -V \int \d x \delta (x) n(x-X)\simeq -V\frac{n_0}{\cosh^2 X/\xi },
\label{5.1}
\eequ
where $X$ is the position of soliton center, $V$ the magnitude of the impurity
potential, $\xi$ is the soliton width, and  $n(x)$ is the charge or spin
density of the soliton. We approximated $n(x)$ as $n_0/\cosh^2  (x/\xi) $
because the electron density in the mid-gap state of the soliton has this
form.\jcite{Takayama}
 Since we are interested in the dynamics of the soliton center and the effect
of soliton trapping.
 We use the effective Hamiltonian.
\bequ
H_{tr}=\frac{P^2}{2 M}-\frac{U}{\cosh^2 X/\xi},
\label{5.2}
\eequ
with $U=|V n_0|$. Here $M$ is the soliton mass which is given by
\bequ
 M=\frac{1}{2\pi\vf} \int\d x \left(\frac{1}{\Del0}\frac{\d \Delta(x)}{\d x}
\right)^2 ,
\label{5.3}
\eequ
and $P$ is the momentum conjugate to the position of the soliton center $X$.
 (\ref{5.2}) gives the trapping energy spectrum given as follows:
\bequ
E_n = -\frac{1}{8 M \xi^2} \left[ \sqrt{1+8MU\xi^2 }-(2 n+1)
\right]^2,
\label{5.4}
\eequ
where $n$ being the non-negative integer which satisfies $0 \leq n \leq
(\sqrt{1+8MU\xi^2}-1)/2 $.
 The experimental value of the soliton formation energy would correspnd to the
sum of
  "pure" formation energy of the soliton $E_s$ and the trapping energy $E_0$.
  The formation energy of the soliton is lowered by
\bequ
E_0 =-\frac{ 1 }{ 8 M\xi^2 }
\left[\sqrt{1+8 M U\xi^2  }-1 \right]^2 .
\label{5.5}
\eequ
In the weak trapping limit, $MU\xi^2<<1$, $E_0$ becomes $- 2MU^2\xi^2 $. When
$MU\xi^2>>1$, $E_0$ becomes $-U+\sqrt{U/(2M\xi^2)}$.

\section{Isotope effect}

The realistic t-polyacetylene has several normal modes of lattice oscillations.
 Horovitz has proposed the model which include these normal modes introducing
plural components of phonon fields.

The  multi-mode Hamiltonian is given by \jcite{Horovits}
\beqarr
H & = & \sum_{\al=1}^3 \frac{1}{2\pi v_F \lambda \omeg0^2} \int \d x
[\dot{\Delta}_\al^2(x)+\omega_\al^2\Delta_\al^2(x) ] \nonumber \\
 & & \nonumber \\
 & & + \sum_s\int\d x \psi_s^\dagger (x)[-\i v_F\sigma_3\partial_x
+ \sigma_1 \Delta(x)] \psi_s(x),
\label{4.1}
\eeqarr
where $3$ types of phonon fields $\Delta_\al$ are taken into account,
$\omega_0$ being the renormalized phonon frequency
$$
\omega_0^2=2\lambda\omega_Q^2.
$$
$\Delta(x)$ is defined by
\bequ
\Delta(x)=\sum_\al (\lambda_\al/\lambda)^{1/2}(\omga/\omeg0) \Delta_\al(x).
\label{4.2}
\eequ
where $\lambda_\al$ is the coupling constant of the $\al$th phonon with the
bare frequency $\omga$.
 There is a relation
\bequ
\lambda=\sum_\al \lambda_\al.
\eequ
 Using (\ref{4.1}), the effective potential is obtained as in the same way as
shown in the section 2.
Introducing a phonon field of the $\alpha$th normal mode by $\eta_\alpha$,
$\Delta$ is expanded around $\Delta (x)$ as follows:
\beqarr
\Delta  & = & \Delta (x) + \eta, \nonumber \\
\eta & = &  \sum_{\alpha} \sqrt{\frac{\lambda_\al}{\lambda}}
\frac{\omega_\alpha}{\omega_0} \eta_\alpha .
\label{4.3}
\eeqarr
Substituting (\ref{4.3}) into the effective Hamiltonian (\ref{1.8}), we obtain
\beqarr
 H & = & -\frac{1}{2\pi\vf} \int \Delta^2(x)\d x
    +
  \frac{ 1}{ 2\pi\vf\lambda\omega_0^2} \sum_\al \int\d x [\dot{\eta}_\al
^2+\omega_\al ^2 \eta^2_\al ]
\nonumber \\
 & & \nonumber \\
& & + \frac{1}{\pi\vf} \sum_{\al , \beta} \int\d x
\left[
 \frac{\sqrt{\lambda_\al \lambda_\beta}}{\lambda}
\frac{\omega_\al \omega_\beta}{\omega_0^2}
  \eta_\al D_\gamma  \eta_\beta
 \right .
\nonumber \\
 & & \nonumber \\
& &
\left. +\left(1+\ln\frac{|\Delta(x)|}{\Delta_0} -\frac{1}{ 2 \lambda }
            \right) \frac{\sqrt{\lambda_\al \lambda_\beta}}{\lambda}
\frac{\omega_\al \omega_\beta}{\omega_0^2}
 \eta_\al \eta_\beta \right] .
\label{4.4}
\eeqarr
where $D_\gamma $ is given by (\ref{di}).
   The eigen mode of the Hamiltonian is constructed as follows.
 From (\ref{4.4}), the Schr{\"o}dinger equation of an eigen mode
$\varphi_\al$ is given by
\beqarr
-\ddot{\varphi}_\al  & = & \omega_\al ^2 \varphi_\al
   +
 2 \sum_\beta
\sqrt{\lambda_\al \lambda_\beta}
\omega_\al \omega_\beta
  D_\gamma \varphi_\beta
\nonumber \\
   & &
     +2\sum_\beta \sqrt{\lambda_\al \lambda_\beta}
\omega_\al \omega_\beta
\left(1+\ln\frac{|\Delta(x)|}{\Delta_0} -\frac{1}{ 2\lambda }
            \right)
  \varphi_\beta .
\label{4.5}
\eeqarr
 Since we are considering the one-dimensional case, the eigen-mode function of
(\ref{4.5}) can be expressed by the super position of $g_\ell$
\bequ
\varphi_\al = \sum_{\ell} C_{\al,\ell} g_\ell ,
\label{4.6}
\eequ
where $g_\ell$ is the eigen function of (\ref{1.9}). We write the frequency of
$g_\ell$ by $\Omega_\ell$. Substituting (\ref{4.6}) into (\ref{4.5}), and using
 eq.(\ref{1.9}),
$C_{\al,\beta\ell}$ is given by the equation,
\bequ
C_{\al,\ell} \sqrt{ \frac{\lambda }{\lambda_\al } }
\omega_\alpha
\left(1- \frac{\Omega^2}{\omega_\al^2} \right) =
\sum_\gamma
\left(1- \frac{\Omega_\ell^2}{\omega_Q^2} \right)
\sqrt{ \frac{\lambda_\gamma }{\lambda } }
\omega_\gamma
C_{\gamma,\ell} .
\label{4.7}
 \eequ
 This equation was obtained in a different way using the RPA method by Terai
{\it et al.}\jcite{Terai}
{}From (\ref{4.7}), the expansion coefficient $ C_{\al,\ell} $
and the frequency $\Omega$ is obtained.
 $\Omega$ is given by the solution of the equation
\beqarr
 \left( 1-\frac{\omega_\ell^2}{\omega_Q^2}  \right)^{-1}
 & = & -D(\Omega),
\nonumber \\
  D(\Omega) & = & \sum_{\al} \frac{\lambda_\al}{\lambda}
 \frac{\omega_\al^2}{\Omega^2-\omega_\al^2}.
\label{4.8}
\eeqarr
There are three solutions for each $\ell$. Let us call them as
$\Omega_{\beta,\ell}$ $(\beta=1,2,3)$. Using (\ref{4.7}) and (\ref{4.8}),
$C_{\al,\ell}$ is given by
\bequ
 C_{\al,\ell} = \frac{\omega_\al}{\omega_{\al}^2-\Omega_{\beta,\ell}^2}
\left( \frac{\lambda_\al}{\lambda} \right) ^{1/2}
\left( \sum_\gamma \frac{\lambda_\gamma}{\lambda}
\frac{\omega_\gamma^2}
{
(\omega_{\gamma}^2-\Omega_{\beta,\ell}^2)^2
}
\right) ^{-1/2}
\equiv
C_{\al,\ell}(\Omega_{\beta,\ell}) .
\label{4.9}
\eequ

 Using (\ref{4.6}), (\ref{4.8}) and(\ref{4.9}), the phonon frequency and the
eigen-mode function are obtained.

Substituting the values in Table II into (\ref{4.8}), the localized modes
around a soliton are obtained. They are summarized in Table I for isotopes of
(CH)$_x$ and (CD)$_x$. The experimental results are also shown.
 In the calculation of the phonon frequency, we used the parameters
$\lambda_\al$ and $\omega_\al$ shown in Table III. They were determined by
Vardeny {\it et al. } from the Raman scattering experiment.
  The phonon frequencies in (CH)$_x$ are relatively
 larger than those in (CD)$_x$.

Let us compare the formation energy of the soliton between the two isotopes of
t-polyacetylene.
As we have discussed in \S 3, the renormalized Hamiltonian is constructed by
adding the counter term, and the soliton formation energy should be corrected
by the fluctuation $\delta m = \left< \eta^2 \right>/\Delta_0^2$.
Using the dispersion relation $\Omega_{\al,k}$ which is obtained from
(\ref{4.8}) for the dimerized state, and using the phonon Green function which
is derived from (\ref{4.4}),
 $\delta m$ is estimated as follows :
\beqarr
\delta m & = &
%\frac{\left< \eta^2 \right>}{\Del0^2} =
\sum_\al \frac{\lambda_\al\omega_\al }{\lambda\omega_{0}
\Del0^2 }
\left<
{\rm T}\tau \ \eta_{\al} (x,\tau=0^+)\eta_{\al}(x,0)
\right> = \frac{\omega_0}{4\Del0 ^2} \sum_\al \int \d (v_F q) \frac{\lambda_\al
\omega_\al}{\Omega_{\al,q}}
 \nonumber \\
 &  = &\left\{ \begin{array}{lr}
 0.4536 & \mbox{for (CH)$_x$} \\
 0.4002 & \mbox{for (CD)$_x$}
         \end{array}\right.
\label{4.10}
\eeqarr
 In the calculation in (\ref{4.10}), we used $\Omega_{\al,q}$ which is given by
(\ref{1.10'}).
 In Appendix B, we have solved (\ref{1.9}) for the soliton approximately.
Since we have made an expansion of $D_\gamma$ up to 6th order with respect to
$\gamma$, we cannot exactly reproduce the dispersion relation (\ref{1.10'}).
 Then, we will use the phonon frequencies given by the RPA method which is
explained in ref.\cite{Ito}. Substituting the phonon frequencies and
(\ref{4.10}) into (\ref{2.9}), the correction to the soliton formation energy
is totally given by %
 \bequ
 \Delta E_s = \left\{ \begin{array}{lr}
       0.0323 \mbox{eV} ,  & \mbox{for (CH)$_x$} \\
       0.0033 \mbox{eV} ,  & \mbox{for (CD)$_x$}.
              \end{array}\right.
\label{4.11}
\eequ
  On the assumption that the difference in the electron band gap $\Delta_0$
between the two isotopes is very small, the corrected parameters are obtained
by (\ref{2.7"}) where the parameters  $\lambda=0.2$ and $\Del0=0.656$eV are
used for both isotopes. The corrected coupling constant $\lambda_R$ for(CH)$_x$
becomes larger than that for (CD)$_x$.
 Then, the soliton formation energy in (CH)$_x$ becomes  about $0.029$eV larger
than that in (CD)$_x$.

The isotope difference of the soliton formation energy is observed by the
photo-induced absorption experiment by Schaffer {\it et
al.}\jcite{Schaffer}\jcite{Heeger}
They concluded that, from the observed value, the difference is about $0.03$eV
$\sim$ $ 0.04$eV.
This is of the same order with our theoretical estimation.
However, the experimental result is a bit larger than ours by about $0.01$eV.
This difference would come from the trapping by impurities.
In the optical absorption experiment, the frequency of $T_1$ mode, which
corresponds to the excitation of soliton translational motion in (CH)$_x$ is
larger than in (CD)$_x$.
This difference is considered to come from the soliton mass $M_s$.
In the multi-component theory (the amplitude mode formalism), the soliton mass
is given by \jcite{Wada}
\bequ
 M_s = \frac{1}{\pi\vf\lambda} \sum_\alpha \frac{\lambda_\al}{
\lambda{\omega_\alpha}^2}
\int\d x \left( \frac{\d \Delta_s (x)}{\d x} \right)^2
=\frac{4 {u_0}^2}{ a} \sum_\alpha \Biggl(
\frac{K}{K_\alpha}  \Biggr)^2 M_\alpha
\int\d x \left( \frac{1}{\Del0} \frac{\d \Delta_s (x)}{\d x} \right)^2
\label{4.12}
\eequ
where $M_\alpha$ is the effective mass of the $\alpha$th lattice oscillation
mode and $K_\alpha$ is the effective spring constant of   the $\alpha$th mode.
They are introduced by the following relation.
\beqarr
\Delta_0 & = & g u_0, \nonumber \\
\lambda & = & \frac{g^2 a}{4\pi v_F K_\alpha}, \nonumber \\
\omega_\alpha & = & \sqrt{ \frac{ K_\alpha }{ M_\alpha } } ,
\eeqarr

  Since the $\sigma$ bonds between atoms get small influences from the hydrogen
mass, the isotope difference of $K_\alpha$ would be small.
  On the other hand, the effective mass of lattice oscillation in (CH)$_x$ is
smaller than that in (CD)$_x$.  Therefore, from (\ref{4.12}), the soliton mass
in (CH)$_x$ is smaller than in (CD)$_x$.
 Thus, in sufficiently deep impurity potential, from (\ref{5.5}), the trapping
energy in (CD)$_x$ is lower than (CH)$_x$, and the excitation energy from the
lowest bound state to higher excited state in (CH)$_x$ is larger than that in
(CD)$_x$.
 In \S 5 we saw that the soliton formation energy is affected by the soliton
trapping energy.
 Suppose that the impurity potential is sufficiently deep and
 that the potential is well-approximated by the harmonic potential
 at the bottom.
 Considering the soliton trapping energy as the energy of the zero-point
oscillation around the impurity potential, the isotope difference of the
trapping energy is about $(0.062-0.0507)/2=0.006$eV, where the experimental
data of $T_1$ mode which is shown in Table I is used.
 Therefore, the soliton formation energy in (CH)$_x$ is larger than that in
(CD)$_x$ by $0.0029+0.006=0.035$eV.
 This theoretical estimate agrees well with the observed value.

\section{Summary}

 We investigated the effect of lattice fluctuation on a soliton and compared
 our result with experimental result observed by the photo-induced optical
absorption experiment.
 First, we renormalized the electron-phonon interaction into the effective
potential of lattice distortion field.
 Then, the effective Hamiltonian is solved to give the soliton solution. We
obtained the dispersion relation of the lattice fluctuation
around the soliton using the effective potential and reproduced the theoretical
result given by using the RPA method.
 There is a good agreement between our result and experimental result about the
level position of the localized mode around the soliton.

 Due to the anharmonicity of lattice vibration, the phonon frequency and the
electron band gap in the actual potential is affected from
 the mean-field value.
We proposed the renormalized Hamiltonian (\ref{2.7'}) and estimated the
correction to the soliton formation energy.
The magnitude of the correction is estimated by the one-loop renormalization
$\delta m=\left<\eta^2\right>/\Delta_0^2$  where $\eta$ is the lattice
fluctuation field and $\Del0$
is the observed value of the electron band gap. We found that the actual depth
of the double-well potential is wider than the depth expected by the mean-field
theory.
The correction to the formation energy due to the electron-phonon coupling is
given by
\bequ
\Delta E_s = \sum_\ell \frac{1}{2}\Omega_\ell\biggl|_{\mbox{soliton}}
 - \sum_\ell \frac{1}{2}\Omega_\ell\biggl|_{\mbox{dimer}}
+\left( (1-\delta m)^{-0.88} -1 \right)
 \frac{\Del0}{\pi} .
\label{6.1}
\eequ
On the assumption that there are differences in the electron-phonon coupling
constant and the phonon frequency between the two isotopes,
 we found that the formation energy in (CH)$_x$ is larger than that in (CD)$_x$
by about $0.029$eV.
  In the realistic material, the soliton formation energy would be affected by
the soliton trapping by the impurities.
  We have shown that the isotope difference of the soliton formation energy is
widened by the trapping energy.
   In the experimental result, the difference between the two isotopes is about
$0.03$eV $\sim$ $0.04$eV.
   The difference from our result is of the same order with the trapping
energy. This suggests the relevance of further theoretical and experimental
investigation of the soliton dynamics.

%\newpage
{\large{\bf acknowledgement }}

Scientific Research from the Ministry of Education, Science, and Culture. The
numerical calculations have been performed on the IBM RISC system 360 of
Department of Physics, Faculty of Education, Yokohama National University.

\appendix
\section{Soliton solution}

 The Hamiltonian (\ref{1.8}) has a kink solution (soliton) connecting the two
minimum points $\Delta = \pm \Del0$.
Let us calculate a static soliton using (\ref{1.8}). The Lagrange equation of
(\ref{1.8}) in the static limit is given by
\bequ
2 \left( \frac{  \sqrt{ 1+\gamma^2 }  }
{\gamma}\ln (\gamma+\sqrt{1+\gamma^2})-1
\right)
\Delta
+2\Delta\ln\frac{ |\Delta| }{2\Lambda}  +\frac{\Delta}{\lambda}=0 ,
\label{a1}
\eequ
where $\gamma=-\i \frac{\xi_0}{2} \frac{\partial }{\partial x}$.  $\xi_0$ is
given by $\xi_0=\vf/\Del0$. Using (\ref{1.7'}) and introducing the notation
$u=\Delta/\Del0$, (\ref{a1}) is rewritten as follows:
\bequ
\frac{1}{6}\xi_0^2\frac{\d^2 u}{\d x^2} = 2u \ln |u| .
\label{a1s}
\eequ
Here we expanded the left hand side of (\ref{a1}) up to the second order with
respect to $\gamma$. The soliton solution is obtained under the condition
\beqarr
  u(x) & = & \pm 1, \quad\quad \mbox{at} (x\rightarrow \pm\infty),
 \nonumber \\
 u(0) & = & 0.
\label{a2}
\eeqarr
Multiplying $\frac{\d u}{\d x}$ to (\ref{a1s}), and making an integration over
$x$, (\ref{a1s}) becomes
\bequ
\frac{1}{ \sqrt{ 6 } }\xi_0\int_0^{x_0}
\frac{ \frac{\partial u}{\partial x} }{\sqrt{ 1+u^2(2\ln |u| -1) }}
\d x = \int_0^{x_0} \d x
\label{a3}
\eequ
{}From (\ref{a3}), the soliton is given by
\beqarr
 \Delta (x) & = & \Del0 u =\Del0 g^{-1} \left(\sqrt{6}
 \frac{x}{\xi_0} \right),
\nonumber \\
 g(u) & = & \int^u_0\frac{\d u}{\sqrt{1+u^2 (2\ln |u| -1)}} .
\label{a4}
\eeqarr
The dotted line in Fig. 4 shows
$\Delta(x)$ given by (\ref{a4}). It is nearly equal to  $\Del0\tanh (2.2
x)/\xi_0$.

 (\ref{a4}) is narrower than $\tanh x/\xi_0$. However, considering the higher
order derivatives in (\ref{a1}) and using the variational method, it is shown
that the soliton width becomes wider and its form is  similar to the form
$\tanh(x/\xi_0)$ .
Since the function (\ref{a4}) would be well approximated by $\sgn
(z)\Del0(1-\exp(-3.88|z|))(1-0.497\exp(-3.88|z|))$, where $z=x/\xi_0$,  we use
the trial function
\bequ
\Delta_s(x)= \sgn (z)\Del0(1-\exp(-a|z|))(1-0.497\exp(-a|z|)) ,
\label{a5}
\eequ
where $a$ is a real number.
Substituting (\ref{a5}) into (\ref{1.8}) and taking the static limit, the mean
value of (\ref{1.8}) is larger than
the dimerized state by
\bequ
\frac{\Del0}{\pi a}\left(
-{\rm Re}\left[
1.249\left(
\frac{\sqrt{4-a^2}}{a} \sin^{-1}\frac{a}{2}-1\right)
+0.1245\left(
\frac{\sqrt{1-a^2}}{a} \sin^{-1}a-1\right)
\right] +{0.9452} \right) ,
\label{a6}
\eequ
where we have omitted the imaginary part of the mean value. We used the
following relation.
\bequ
\int\Delta(x) \frac{\partial ^{2n}}{\partial ^{2n} z} \Delta(x)
\d z =-2a^{2n-1} [0.6305  +0.12351(2)^{2n-1}] .
\label{a6s}
\eequ
 (\ref{a6}) has a minimum point at $a=1.70$.
 Then, (\ref{a5}) is $2.28$ times wider than the function (\ref{a4}).
  Solid line in Fig.4 shows (\ref{a5}) with $a=1.70$.
  Its form is nearly equal to the form of
  \bequ
  \Delta_s(x)=\Del0\tanh (x/\xi) ,
  \label{a7"}
  \eequ
  with
\bequ
 \xi = 1.04 \xi_0 .
 \label{a7}
\eequ

Using that $\Delta(x)$ is the solution of (\ref{a1}), the soliton formation
energy is given by
\bequ
\frac{1}{2\pi v_F}\int \d x ({\Del0}^2-\Delta_s^2(x) ).
\label{a7'}
\eequ
  Since the soliton form  (\ref{a7"}) has the same form as the solution of the
original Hamiltonian (\ref{1.1}), it is adequate to use $\Delta_s (x) =
\Del0\tanh (x/\xi_0)$ for the estimation of (\ref{a7'}). Therefore, (\ref{a7'})
becomes $\Del0/\pi$.

 The correction due to the renormalization of the unharmonic potential gives
the effective Hamiltonian (\ref{2.7}). When $d_R$ is varied, the soliton width
$\xi_R$, which is obtained by using the trial function (\ref{a5}), changes. We
plotted $\xi_R$ for various values of ${d_R}^{-1}$ in Fig.7.

\section{Lattice fluctuation around a soliton}

Differentiating (\ref{a1}) with respect to $x/\xi_0=\kappa x$, we obtain the
following equation.
\bequ
 \left( \frac{  \sqrt{ 1+\gamma^2 }  }
{\gamma}\ln (\gamma+\sqrt{1+\gamma^2})-1
\right)
\frac{\d \Delta_s (x)}{\d (\kappa x) }
+\left(1+\ln\frac{ |\Delta_s (x)| }{|\Del0|}\right)
\frac{\d \Delta_s (x)}{\d (\kappa x)} =0.
\label{b1}
\eequ
Comparing (\ref{b1}) with (\ref{1.9}), we find that $\frac{\d \Delta(x)}{\d
(\kappa x)}$ is an eigen mode with zero energy, {\it i.e.} the Goldstone mode.

In the calculation of the other eigen mode around the soliton, we use the
soliton form obtained in Appendix A.
The eigen-mode equation (\ref{1.8}) is approximated by
\bequ
-\frac{1}{2\lambda\omega_Q^2} \ddot{g}_\ell
 =  -  \left( \frac{1}{12} \frac{\partial^2}{\partial (\kappa x)^2 } +
  \frac{1}{120} \frac{\partial^4}{\partial (\kappa x)^4 } +
  \frac{1}{840} \frac{\partial^6}{\partial (\kappa x)^6 }
\right) g_\ell
 + \left( 1 +\ln \frac{|\Delta(x)|}{\Del0} \right) g_\ell .
\label{b2}
\eequ
where we expanded $D_\gamma$ until 6th derivatives with respect to $\kappa x$.
Using (\ref{a5}) with $a=1.70$, the eigen mode is determined by the numerical
diagonalization.  There are two additional modes with localized character.  The
eigen values of the localized modes are summarized in Table I.
 They have a good coincidence with the result given by ref.\cite{LM}.

\begin{figcaption}

\item  The loop diagrams which contribute to the effective potential of the
lattice distortion.
These diagrams do not have the momentum transfer between the electron and the
lattice.

\item  The loop diagrams which contribute to the effective potential of the
lattice distortion.
These diagrams have the  momentum transfer $q$ between the electron and the
lattice.

\item The $Ê\Delta$ dependence of the effective potential. The dotted line is
the effective potential which is given by the Hamiltonian (\ref{1.8}). The
solid line is the potential  (\ref{2.1})

\item  The soliton configuration. The dotted line is the solution of the
soliton equation which is approximated by the second order differential
equation (\ref{a1s}). The solid line is the soliton which is given by the
variational method. This form is nearly equal to the form $\tanh(x/\xi_R)$.

\item  The diagrams of the 3 point and 4 point vertices which appear in the
Hamiltonian (\ref{2.1}).

\item  The one loop renormalization diagrams for the 3 point vertex and 4 point
vertex, respectively.

\item The ${d_R}^{-1}$ dependence of soliton width. The soliton is obtained by
the variational method using the renormalized Hamiltonian (\ref{2.7'}), where
the trial function is chosen to be (\ref{a5}). Since the function (\ref{a5})
has the form nearly equal to $\tanh (0.567 a x/\xi_0)$, the soliton width
$\xi_R$ is given by $\xi_R=\xi_0/(0.567a)$.
\end{figcaption}

\noindent Table I: The frequencies of the localized modes around the soliton.
The T modes correspond to the translational mode of soliton and the B modes
come from the third localized mode around the soliton.

\vspace{12pt}

\noindent Table II :
We summarized the values of $\omega^2/(2\lambda{\omega_Q}^2)$ for the localized
modes of the lattice fluctuations around the soliton. They are obtained by the
numerical diagonalization of (\ref{1.8}) which is approximated by the
differential equation (\ref{b2}).

\vspace{12pt}

\noindent Table III :  The bare values of the phonon frequencies and the
electron-phonon coupling constants.  These values were used in the analysis of
the Raman scattering experiment of t-polyacetylene by Vardeny {\rm et.
al.}\jcite{VS}

\newpage
\begin{center}

\begin{tabular}{ r c c c c c c} \hline
            & \multicolumn{2}{c}{Theory$^{1)}$}  &
 \multicolumn{2}{c}{Present Result}  & \multicolumn{2}{c}{Experiment} \\
\hline
            &     $(CH)_x$    &  $(CD)_x$  & $(CH)_x$ &  $(CD)_x$ & $(CH)_x$ &
$(CD)_x$  \\   \hline
   $ T_1\ (\omega_{00})$   &    488         &  410   & 0.     & 0.      &
$\sim$ 500 & $\sim$ 400 \\
   $ T_2\ (\omega_{10})$   &    1278      & 1045  & 1270 & 1014 & 1280  & 1045
 \\
   $ T_3\ (\omega_{20})$   &    1364       & 1216  & 1344 & 1207& 1365  & 1224
 \\   \hline
   $ B_1\ (\omega_{02})$   &    1049       & 850  &1079 & 862 & 1035  & 835
 \\
   $ B_2\ (\omega_{12})$   &    1292      & 1193  & 1293 & 1197 &  -      & -
 \\
   $ B_3\ (\omega_{22})$   &    1445       & 1333  & 1463 & 1368  &1440  & 1300
 \\   \hline
\end{tabular}

%\begin{tabular}{ r c c c c } \hline
%            & \multicolumn{2}{c}{Theory}  & %\multicolumn{2}{c}{Experiment} \\
%%\hline
%            &     $(CH)_x$    &  $(CD)_x$  & $(CH)_x$ &  $(CD)_x$  \\
%%%\hline
%   $ T_1\ (\omega_{00})$   &     488          &  410        & $\sim$ %500   &
%%$\sim$ 400     \\
%   $ T_2\ (\omega_{10})$   &    1278       & 1045     & 1280  & %1045     \\
%   $ T_3\ (\omega_{20})$   &    1364       & 1216     & 1365  & %1224     \\
%%\hline
%   $ S_1\ (\omega_{01})$   &    0.1677       & 0.1468     &  -      & -
%%  %\\
%   $ S_2\ (\omega_{11})$   &    0.2432       & 0.2169     &  -      & -
%%  %\\
%   $ S_3\ (\omega_{21})$   &    0.2635       & 0.2341     &  -      & -
%%  %\\   \hline
%   $ B_1\ (\omega_{02})$   &    1079       & 862     & 1035  & 835     %\\
%   $ B_2\ (\omega_{12})$   &    1293       & 1197     &  -      & -
%%%\\
%   $ B_3\ (\omega_{22})$   &    1463       & 1368     & 1440  & %1300     \\
%%\hline
%
%\end{tabular}

\vspace{3cm}

\begin{tabular}{ r c c c } \hline
           &    Goldstone mode   &  Amplitude oscillation mode & Third
localized mode   \\   \hline
  $  \omega^2/ (2\lambda \omega_Q^ 2 )  $   &    0.00       & 0.7186    &
0.9593   \\   \hline
\end{tabular}

%\newpage
\vspace{3cm}

\begin{tabular}{ r c c c c } \hline
            & \multicolumn{2}{c}{(CH)$_x$}  & \multicolumn{2}{c}{(CD)$_x$} \\
\hline
            &  $\lambda_\alpha/\lambda$    &  $\omega_\alpha$  (cm$^{-1}$)
            &  $\lambda_\alpha/\lambda$    &  $\omega_\alpha$  (cm$^{-1}$)  \\
 \hline
   $ \alpha=1$   &     0.07      &  1234     & 0.06   & 921     \\
   $ \alpha=2$   &    0.02       &  1309     & 0.005  & 1207    \\
   $ \alpha=3$   &    0.91       &  2040     & 0.93  &  2040    \\   \hline
\end{tabular}

\end{center}

    %%%%%%%%%%%%%%%%%%%%%%%%%

\end{document}